\documentclass[twocolumn,prl,aps,showpacs,superscriptaddress]{revtex4}
\usepackage{graphicx}
\usepackage{dcolumn}
\usepackage{bm}
\usepackage{epsfig}
\usepackage{pictex}
\begin{document}
\def \Nax{Na$_x$CoO$_2$~}
\def \Naxn{Na$_x$CoO$_2$}
\def \Na{Na$_{0.5}$CoO$_2$~}
\def \Nan{Na$_{0.5}$CoO$_2$}
\def \Kx{K$_x$CoO$_2$~}
\def \Kxn{K$_x$CoO$_2$}
\def \K{K$_{0.5}$CoO$_2$~}
\def \Kn{K$_{0.5}$CoO$_2$}
\def \Rx{$A_x$CoO$_2$~}
\def \Rxn{$A_x$CoO$_2$}
\def \R{$A_{0.5}$CoO$_2$~}
\def \Rn{$A_{0.5}$CoO$_2$}
\newcommand {\ibid}{{\it ibid}. }
\newcommand {\etal}{{\it et al}. }
\newcommand {\etaln}{{\it et al}.}
\newcommand {\etalc}{{\it et al}., }

\title{
Interplay of frustration, magnetism, charge ordering, and covalency in the ionic Hubbard model on the triangular lattice at three-quarters filling}
\author{Jaime Merino}
\affiliation{Departamento de F\'isica Te\'orica de la Materia
Condensada,
Universidad Aut\'onoma de Madrid, Madrid 28049, Spain}
\author{B. J. Powell}
\author{Ross H. McKenzie}
\affiliation{Centre for Organic Photonics and Electronics, School of Physical Sciences, University of Queensland, Brisbane 4072,
Australia}
\date{\today}
\begin{abstract}
We investigate the ionic Hubbard model on a triangular lattice at three-quarters filling.
This model displays a subtle interplay between metallic and insulating phases and between 
charge and magnetic order.
We find crossovers between Mott, charge transfer and covalent insulators and magnetic order with 
large moments that persist even when the charge transfer is weak.
We discuss our findings in the context of recent experiments on the
layered cobaltates A$_{0.5}$CoO$_2$ (A=K, Na).
\end{abstract}
\maketitle


The competition between metallic and insulating states in strongly
correlated materials leads to many novel behaviours. The  Mott
insulator occurs when a single band is half-filled and the on-site
Coulomb repulsion, $U$, is much larger than hopping integral, $t$. A
menagerie of strongly correlated states is found when a system is
driven away from the Mott insulating state, either by doping, as in
the cuprates \cite{Anderson-RVB}, or reducing $U/t$, as in the
organics \cite{organics-review}. Geometric frustration causes yet
more novel physics in Mott systems \cite{organics-review}. Therefore
the observation of strongly correlated phases in the triangular
lattice compounds A$_{0.5}$CoO$_2$,  where $A$ is K or Na \cite{ong-cava-science},
 has created intense interest.

An important model for investigating insulating states in correlated
materials is the ionic Hubbard model. 
On a half-filled square lattice this model displays a          crossover
 between  Mott and band insulating states which 
has been analyzed with quantum Monte Carlo (QMC) \cite{bouadim},  dynamical 
mean field theory (DMFT) and its cluster extensions \cite{kancharla}. 
However, except for the case of one dimension \cite{penc2},
this model has not been studied away from half-filling \cite{penc1} 
and/or on geometrically frustrated lattices.

In this Letter we study the ionic Hubbard model on a triangular lattice at three-quarter
filling. This Hamiltonian displays a subtle interplay between
metallic and insulating phases and charge and magnetic order. It has
regimes analogous to Mott, charge transfer  \cite{Zaanen}, and
covalent insulators \cite{Sarma}. The study of this model is
motivated in part by our recent proposal \cite{MPM} that it is an
effective low-energy Hamiltonian for \Naxn, at values
of $x$ at which ordering of the sodium ions occurs.

The Hamiltonian for the ionic Hubbard model is
\begin{equation}
H=-t\sum_{\langle ij\rangle\sigma} c^\dagger_{i\sigma} c_{j\sigma}  +U \sum_i n_{i \uparrow} n_{i \downarrow}
+\sum_{i\sigma} \epsilon_i n_{i \sigma}, \label{ham}
\label{model}
\end{equation}
where $c^{(\dagger)}_{i\sigma}$ anihilates (creates an electron
with spin $\sigma$ at site $i$, $t$ is the hopping integral, $U$ is
the effective Coulomb repulsion between electrons on the same site,
and $\epsilon_i$ is a the site energy. We specialise to the case
with  two sublattices, A ($\epsilon_i=\Delta/2$) and B
($\epsilon_i=-\Delta/2$), consisting of alternating rows, with
different site energies on the two sublattices (c.f., Fig. 15 of
Ref. \onlinecite{MPM}). This is the lattice relevant to \R where the
difference in site energies results from the ordering of the A-atoms \cite{williams-argyriou,Na_ordering,Na-expt}.

Two limits of model (\ref{model}) at $3/4$-filling 
may be easily understood.
For non-interacting electrons, $U=0$, a metallic 
state occurs  for all $\Delta$ as at least one-band crosses
the Fermi energy. In the atomic limit $t=0$, and 
$U>\Delta$ one expects a charge transfer insulator with a
charge gap of about $\Delta$ whereas for 
$U<\Delta$ a Mott insulator with charge gap of $U$ occurs. 
However, realistic parametrization of \Rx materials
imply $U\gg\Delta$ and $\Delta \sim |t|$ \cite{foot-parms}; we will show below that in this parameter regime the model show very different behaviour from either of the limits discussed above. This interesting regime needs to be analyzed
using non-perturbative and/or numerical techniques.
Thus, we have performed Lanczos diagonalization calculations on 
18 site clusters with periodic boundary conditions.

\begin{figure}
\begin{center}
\epsfig{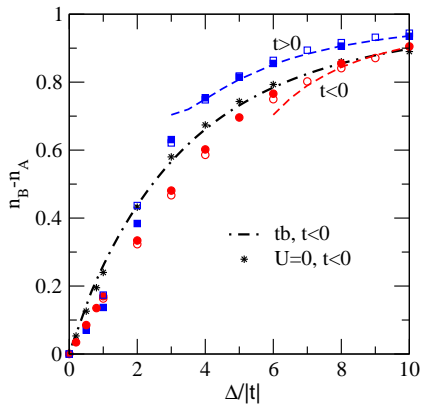}\\
\caption{\label{fig:nanb}
(Color online) Variation of the charge transfer with $\Delta$ for
different Coulomb repulsion energies. 
Lanczos results on an 18-site cluster for: $U/|t|=0$ (stars), 15 (open symbols) and 
100 (filled symbols) are shown. The $U=0$ results are in excellent agreement
with the infinite lattice tight-binding (tb) result (dashed-dotted line) \cite{footnon}.
Strong coupling results (dashed lines) are also shown for comparison. 
}
\end{center}
\end{figure}

In Fig. \ref{fig:nanb} we plot the charge transfer, $n_B-n_A$ as a
function of $\Delta/|t|$ for several values of $U$. We also plot 
$n_B-n_A$ in two analytically tractable limits: the
non-interacting limit, $U=0$ \cite{footnon}; and the strong coupling
limit $U\gg\Delta\gg|t|$ \cite{footstrong}. Several interesting effects
can be observed in this calculation. Firstly, the sign of $t$
strongly effects the degree of charge transfer on the triangular
lattice. Secondly, charge transfer depends only weakly on $U$. Thirdly, regardless of the sign of $t$ or the
magnitude of $U$, the charge transfer  increases rather slowly as
$\Delta$ increases.

\begin{figure}
\begin{center}
\epsfig{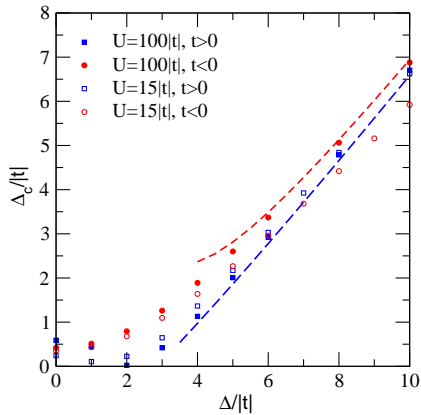}\\
\end{center}
\caption{\label{fig:gap} (Color online) Variation of the charge gap,
$\Delta_c$, with $\Delta$ for $U/|t|=15$ and 100 (the charge gap is
zero for $U=0$). The tendency towards an insulating
state is greater for $t<0$ than for $t>0$. Dashed lines show
the strong coupling limit: $U>>\Delta>>t$. 
Note that although strong correlations are essential for creation of
the charge gap they are not required for the charge transfer, c.f.. Fig. \ref{fig:nanb}. Note that, 
for $U>>\Delta$ with $U$ large, the charge gap is robust against the value of $U$. 
}
\end{figure}

The charge gap, i.e., the difference in the chemical potentials for
electrons and holes, is  $\Delta_c \equiv
E_0(N+1)+E_0(N-1)-2E_0(N)$, where $E_0(N)$ is the ground state
energy for $N$ electrons. We plot the variation of $\Delta_c$ with
$\Delta$ for various values of $U$ in Fig. \ref{fig:gap}. $\Delta_c$
vanishes for $U=0$, however finite size effects mean that we cannot accurately calculate $\Delta_c$ for small $\Delta$.
 $\Delta_c =\Delta$ for $t=0$ and $U \gg
\Delta$; this result is reminiscent of a charge transfer
insulator \cite{Zaanen}.  Both  perturbative \cite{footstronggap}
and numerical results show that the charge gap depends on the sign
of $t$ due to the different magnetic and electronic properties arising from the geometrical
frustration of the triangular lattice. In contrast, on a square lattice, 
$\Delta_c$, does not depend on the sign of $t$.

In the limit, $\Delta\gg U\gg|t|$, the A and B sublattices are well
separated in energy; the B sites are doubly occupied (i.e., the
B-sublattice is a band insulator) and the A sublattice is
half-filled and hence becomes a Mott insulator. 
If there were no hybridisation between that chains, one would find a metallic state for any finite
charge transfer from the B-sites to the A-sites (self doping), even
for $U\gg |t|$ as the A-chains are now electron-doped Mott
insulators and the B-chains are hole-doped band insulators.
However, Fig.
\ref{fig:gap} shows that the insulating regime of the model extends
far beyond the well understood $n_B-n_A=1$ regime.   
This is because the real space interpretation is incorrect as hybridization
between A and B chains is substantial.           
For $|t|\sim\Delta\ll U$ the system can remain
insulating with a small gap [${\cal O}(t)$]. This state is analogous
to a covalent insulator \cite{Sarma}. 

One expects that for $\Delta=0$ the ground state is metallic as there
the system is $3/4$-filled. However, a small but finite $\Delta=0^+$ leads to 
a strongly nested Fermi surface for $t>0$ whereas for $t<0$ the Fermi surface
rather featureless. Thus, rather different behaviors might be expected for different signs of $t$
even at weak  coupling. 
At large $U$ our exact diagonalization results suggest that 
a gap may be present even for a small value of $\Delta/t$.
However finite-size effects, inherent to the method, mean that
it is not possible to resolve whether a gap opens at $\Delta=0$ or at some
finite value of $\Delta$.

\begin{figure}
\begin{center}
\epsfig{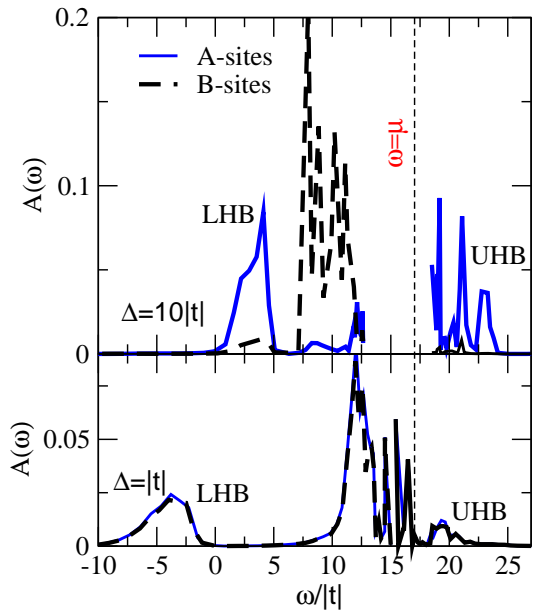}
\end{center}
\caption{\label{fig:dos}(Color online)
Evolution from a charge transfer insulator
to a covalent insulator. The energy dependence of the
spectral density, $A(\omega)$, is shown
for two different parameter regimes of the model with
$t<0$ and $U=15|t|$.
The spectral density in the upper panel ($\Delta=10|t|$)
is that characteristic of a charge transfer insulator \cite{Sarma}
there is a single weakly correlated band largely associated with
the B sites and lying between lower (LHB) and upper (UHB)
Hubbard bands (separated by $\sim U$)
that are largely associated with the A sites and $\Delta_c\sim\Delta$.
In contrast, the lower panel ($\Delta=|t|$) shows a spectral density
characteristic of a covalent insulator \cite{Sarma}:
there are only small difference between A and B sites,
the separation of the LHB and UHB is
 $>U$,
and $\Delta_c\sim |t|$.
}
\end{figure}

To test this covalent insulator interpretation in the $\Delta\sim|t|$ and large $U$
regime we have also calculated the spectral
density, $A(\omega)$, c.f., Fig.  \ref{fig:dos}.  There are three distinct contributions to the
$A(\omega)$: at low energies there is a lower Hubbard band; just
below the chemical potential ($\omega=\mu$) is a weakly correlated
band; and just above $\omega=\mu$ is the upper Hubbard band. Furthermore, the 
large energy separation, much larger than the expected $U=15|t|$, between the 
lower and upper Hubbard bands is due to an upward (downward) shift of the 
upper (lower) Hubbard bands due to the strong hybridization. In contrast, in 
the strong coupling limit  $A(\omega)$  has a much larger gap, ${\cal O}(\Delta)$, 
between the contributions from the weakly correlated band and the upper Hubbard band.

\begin{figure}
\begin{center}
\epsfig{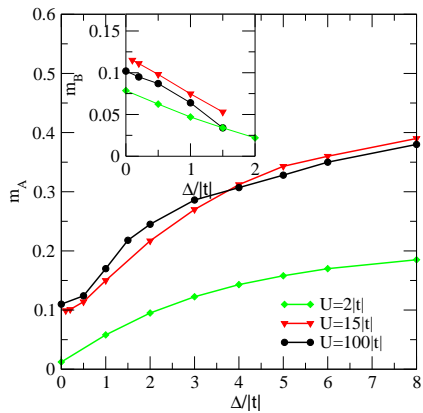}
\end{center}
\caption{\label{fig4}(Color online) The magnetic moment as a
function of $\Delta$ for $t<0$ and various $U$. The magnetic moment
on the A sites, $m_A$, (main panel) is strongly enhanced when $U\gg
|t|$. The inset shows the moment on the B sublattice. $m_B$ is much
smaller,  reduced by $\Delta$, and  only weakly dependent on $U$.
These results demonstrate that, for large $U$ and
$\Delta\gtrsim|t|$, the electrons on the A-sublattice are much more
strongly correlated than those on the B-sublattice despite the small
charge transfer (see Fig. \ref{fig:nanb}). }
\end{figure}

The magnetic moment associated with the possible antiferromagnetism,
$m_ \nu =(3 \langle S^z_{i} S^z_{j} \rangle)^{1/2}$, where $\nu =A$
or  $B$ and $S^z_{i}={1 \over 2} (n_{i \uparrow} -n_{i
\downarrow})$, is  evaluated between two next-nearest neighbors on
the $\nu$ sublattice at the center of cluster (to reduce finite size
effects \cite{sandvik}). Fig. \ref{fig4} shows that $m_A$ increases
with $\Delta$ and is substantially enhanced by $U$, whereas $m_B$ is
always small. This is in marked contrast to a spin density wave, as
predicted by Hartree-Fock calculations where the magnetic
moment is far smaller than that experimentally observed \cite{RPA}.

We now turn to discuss the consequences of our results
for understanding experiments;
for simplicity and concreteness we focus on \Naxn. 
The $x=0.5$ materials
have remarkably different properties from those on other
values of $x$ \cite{foo,wang}. Above 51~K the intralayer resistivity of \Na is
weakly temperature dependent with values of a few m$\Omega$cm
\cite{foo} characteristic of a bad metal \cite{MPM}. Below 51~K the
resistivity increases, consistent with a small gap opening
($\sim$10~meV) \cite{foo}. Thus a (bad) metal-insulator transition
occurs at 51~K. The insulating state of \Na has a number of
counterintuitive properties, not the least of which is the absence
of strong charge ordering. NMR observes no charge ordering up to a
resolution of $n_B-n_A<0.4$ \cite{bobroff,nmr}, while  neutron
crystallography suggest $n_B-n_A\simeq0.12$
\cite{williams-argyriou}. Thus the insulating state is not the
simple charge-transfer-like state predicted by (\ref{model}) in the
strong coupling limit. \Na develops a commensurate magnetic order
 below 88~K \cite{bobroff,gasparovic-yokoi}.
 A large magnetic moment
[$m=0.26(2)\mu_B$ per magnetic Co ion] is observed in spite of the
weak charge order [note that classically $m<(n_B-n_A)\mu_B/2$].
Above 100~K the optical conductivity \cite{wang} shows no evidence
of a Drude peak, consistent with a bad metal. In the insulating
phase spectral weight is lost below $\sim$10~meV, consistent with
the gap seen in the dc conductivity and a  peak emerges at
$\sim$20~meV, which is too sharp and too low energy to correspond to
a Hubbard band  \cite{wang}. ARPES shows that the highest energy
occupied states are $\sim$10~meV below the Fermi energy \cite{qian}.
No equivalent insulating state is seen in the misfit cobaltates
\cite{bobroff}, which supports the contention that Na-ordering is
vital for understanding the insulating state.

Various theories have been proposed to explain these intriguing experiments. Lee \etal \cite{Lee} have performed LDA+U calculations, which include Na-ordering, but not strong correlations. Other groups \cite{theory} have studied strongly correlated models that include the Coulomb interaction with neighbouring sites, but neglect the effects of Na-ordering.   
Marianetti and Kotliar \cite{Marianetti} have also studied the Hamiltonians proposed in \cite{MPM} for $x=0.3$ and 0.7.

In order to compare our results with experiments on \Na we need to
restrict ourselves to the relevant parameter values: $t<0$ and
$|t|\sim\Delta\ll U$ \cite{foot-parms}. This corresponds with the
regime of the three quarters filled ionic Hubbard model that is both
the most interesting and the most difficult to study via exact
diagonalisation because of the deleterious finite size effects.
Nevertheless we propose that in \Na the insulating state is
analogous to a covalent insulator. This explains a wide range of
experiments.  The peak
observed at $\omega \sim30$ meV in the optical conductivity
\cite{wang}, is interpreted as the transfer of an electron from the
weakly correlated band to form a doublon in the strongly correlated
band. The weak charge transfer ($n_B-n_A=0.1-0.3$;
c.f., Fig. \ref{fig:nanb}) is caused by the strong hybridisation
between the A and B sublattices and is consistent with the value
extracted from crystallographic experiments (0.12
\cite{williams-argyriou}) and the bounds from NMR ($<$0.4
\cite{bobroff}). The large moment (0.1-0.2$\mu_B$; c.f., Fig.
\ref{fig4}) is comparable to the moment found by neutron scattering
($0.26\mu_B$ \cite{gasparovic-yokoi}) and results from the electrons
in the strongly correlated band, i.e., the single spin hybridised
between the A and B sublattices.
Finite size effects mean that we cannot accurately calculate the charge gap in this regime. However, we propose that the experimental system corresponds to a parameter range where the gap is small,
$\Delta_c<{\cal O}(|t|)$, consistent with the
gap, $\sim$7-10~meV \cite{foo,qian}, seen experimentally in ARPES
and resistivity. This is consistent with the expectation that $\Delta_c\rightarrow0$ as 
$\Delta/|t|\rightarrow0$.
Accurately calculating $\Delta_c$ for small $\Delta/|t|$ and large $U$, and hence further testing our hypothesis, therefore remains an important theoretical challenge.

\acknowledgments We thank H. Alloul, Y.S. Lee, and R.
 Singh for helpful discussions. J.M. acknowledges financial support from
the Ram\'on y Cajal program, MEC (CTQ2005-09385-C03-03).
 B.J.P. was the recipient  of  an ARC Queen
Elizabeth II Fellowship (DP0878523). R.H.M. was the recipient  of  an ARC Professorial Fellowship (DP0877875). Some of the numerics were performed on the  APAC national facility.


\begin{thebibliography}{999}
\bibitem{foot-parms}
Atomistic calculations give
$t$=-(0.08-0.14) eV,
$U$=2.5-2.8 eV, and
$\Delta$=0.03-0.16 eV  \cite{Na_ordering,singh}.
Further, CoO$_2$, which
is described by
model (\ref{model}) at half filling
with $\Delta=0$,
is a strongly correlated metal rather than a Mott insulator \cite{vaulx}.
This suggests $U/|t| \lesssim 12-15$, the critical
value for a Mott insulator
on the triangular lattice \cite{Capone,MPM}.
\bibitem{footnon}
The tight-binding band structure of model (\ref{model}) consists of two bands due to
hybridization between $A$ and $B$-sites.
At $3/4$-filling, for $t>0$ and any non-zero $\Delta$,
  the lower band is filled
 and the upper band is half-filled, while for $t<0$ the lower band is filled for $|\Delta|\gtrsim0.64t$.
\bibitem{footstronggap}
The lowest order correction to the charge
gap, in the strong-coupling limit, comes from the kinetic energy of
a hole (doublon) propagating along the B(A)-chains
 when extracting (adding) an electron
to the zeroth order ground state configuration. Using degenerate
perturbation theory we find that the gap is given,
 to O($t^3$), by
$ \Delta_c = \Delta-2|t|+4t^2/\Delta-8t^3/\Delta^2 +E_{1D}^{t-J} $,
where $E_{1D}^{t-J}$ is the energy of a hole in a half-filled
$t$-$J$ chain with $J$ given by the expression in Ref.
\onlinecite{footstrong}.
$E_{1D}^{t-J}$ is given by the Bethe ansatz expression \cite{Ogata}
for $J>0$, while $E_{1D}^{t-J}=-2|t|-J/2$ for $J<0$.
\end{thebibliography}
\end{document}